\DocumentMetadata{}
\documentclass[sigconf,nonacm]{acmart}

\usepackage{xspace}
\usepackage{balance}
\usepackage{comment}
\usepackage{enumitem}
\usepackage{xcolor}
\usepackage{soul}

\def\etal{\textit{et~al.}\xspace}

\begin{document}

\def\sysname{\mbox{MemoryLane}}

\author{Soumay Agarwal}
\orcid{0009-0009-4127-5610}
\email{soagarwa@fiu.edu}
\affiliation{%
  \institution{Florida International University}
  \city{Miami}
  \state{FL}
  \country{USA}
}

\author{Aaryan Gajula}
\orcid{0009-0002-2846-2007}
\email{agaju003@fiu.edu}
\affiliation{%
  \institution{Florida International University}
  \city{Miami}
  \state{FL}
  \country{USA}
}

\author{Lingyao Li}
\orcid{0000-0001-5888-8311}
\email{lingyaoli@arizona.edu}
\affiliation{%
  \institution{University of Arizona}
  \city{Tucson}
  \state{AZ}
  \country{USA}
}

\author{Renkai Ma}
\orcid{0000-0002-4434-2235}
\email{mark@ucmail.uc.edu}
\affiliation{%
  \institution{University of Cincinnati}
  \city{Cincinnati}
  \state{OH}
  \country{USA}
}

\author{Jennifer Martin}
\orcid{0000-0003-0849-3391}
\email{jlmartin@fiu.edu}
\affiliation{%
  \institution{Florida International University and Miami VA Healthcare System}
  \city{Miami}
  \state{FL}
  \country{USA}
}

\author{Krisstina Madan}
\orcid{0009-0008-1229-5105}
\email{kmadan@fiu.edu}
\affiliation{%
  \institution{Florida International University}
  \city{Miami}
  \state{FL}
  \country{USA}
}

\author{Ellen Brown}
\orcid{0000-0002-2418-3257}
\email{ebrown@fiu.edu}
\affiliation{%
  \institution{Florida International University}
  \city{Miami}
  \state{FL}
  \country{USA}
}

\author{Chen Chen}
\orcid{0000-0001-7179-0861}
\email{chechen@fiu.edu}
\affiliation{%
  \institution{Florida International University}
  \city{Miami}
  \state{FL}
  \country{USA}
}

\renewcommand{\shortauthors}{Agarwal~\etal}


\title[\sysname]{\sysname: Supporting Reminiscence Conversations Between Older Adults with Mild Dementia and Lay Caregivers Using a Simulated AI Agent}

\begin{teaserfigure}
    \centering
    \includegraphics[width=\linewidth]{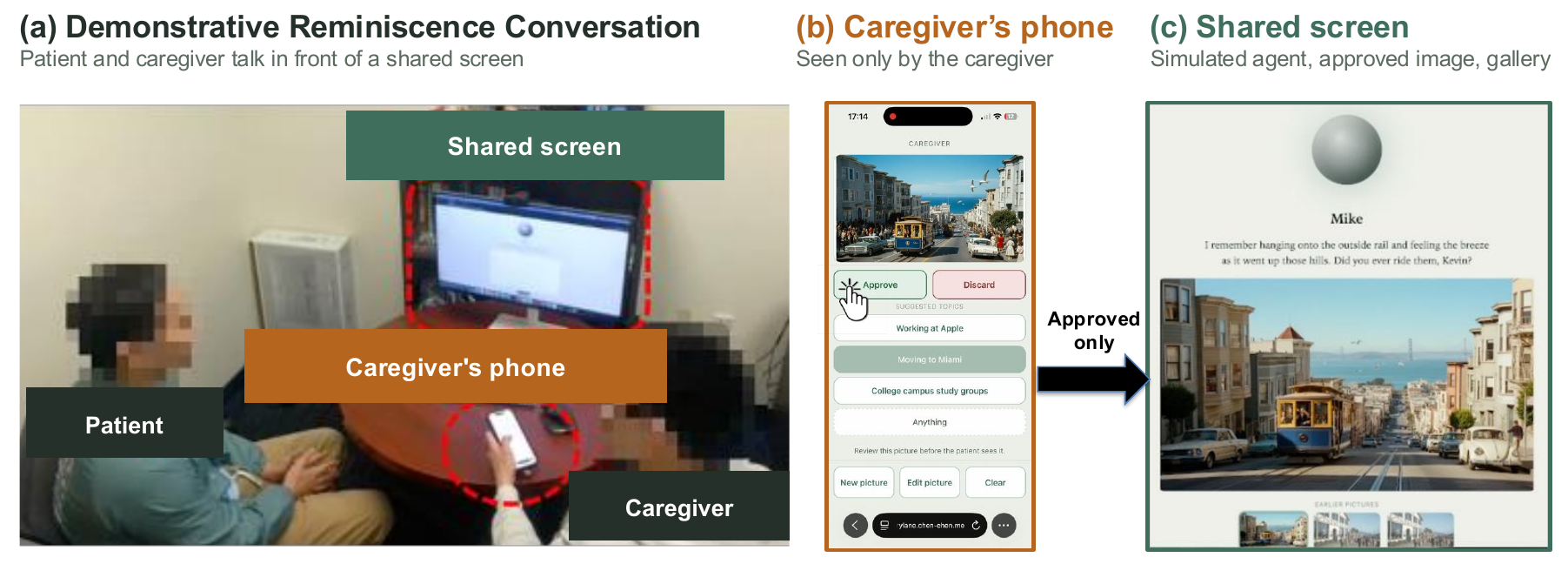}
    \vspace{-.25in}
\caption{(a) The patient and caregiver talk in front of a shared screen while the caregiver holds a phone. (b) On the private phone, the caregiver reviews each generated image, picks from suggested topics or invites the companion to speak, and requests, edits, or clears images. (c) Only approved images reach the shared screen, alongside the AI companion and a gallery of earlier images.}
    \label{fig::teaser}
    \Description{Three-panel figure. Panel (a) is a photo of an older adult and a caregiver seated at a table in front of a computer monitor, with the caregiver holding a smartphone; labels mark the patient, caregiver, shared screen, and caregiver's phone. Panel (b) is a screenshot of the caregiver's phone interface showing a generated street scene with a cable car awaiting review, Approve and Discard buttons, a list of suggested conversation topics with an "Anything" option, and New picture, Edit picture, and Clear buttons. An arrow labeled "Approved only" points from the phone to panel (c), a screenshot of the shared screen showing the AI agent's animated orb, its spoken caption, the approved cable car image, and a gallery of earlier images.}
\end{teaserfigure}

\begin{abstract}
Reminiscence Therapy (RT) is a non-pharmacological intervention that invites patients to discuss and review past life experiences and memories to improve their emotional and mental well-being. In practice, many reminiscence conversations are led by lay caregivers, such as spouses and adult children. While traditional RT often relies on personal photographs, curating them can be tedious and difficult. In addition, caregivers who lack shared life experiences with the patient may struggle to facilitate engaging reminiscence conversations.
This paper presents \textit{MemoryLane}, an AI-integrated tool that helps lay caregivers facilitate reminiscence conversations without curated photographs. MemoryLane supports a triadic conversation among the patient, the caregiver, and a simulated AI agent that acts as the patient's peer. A shared interface shows the agent and AI-generated images of the memories being discussed, while a private smartphone interface lets the caregiver control the behavior of the simulated AI agent and moderate the AI-generated images. A preliminary focus group study with five dementia-care stakeholders suggested the value of MemoryLane's AI agent as a peer companion and of AI-generated images, as well as the feasibility of having caregivers moderate images and control the AI agent.
\end{abstract}

\maketitle

\section{Introduction}

Population aging and the rising prevalence of Alzheimer’s Disease and Related Dementias (ADRD) are a growing global concern, creating an urgent need for effective interventions to support patients at every stage of ADRD~\cite{Altizer2025}.
Reminiscence Therapy (RT) is a non-pharmacological intervention designed to invite patients to discuss and review past life experiences, events, and memories to improve emotional and mental well-being.
While traditional RT often relies on personal photographs, many patients with ADRD may lack such photos or have difficulty accessing them, particularly immigrants and those of low socioeconomic status~\cite{Nan2025Kimono, Zhang2025RemiHaven}.
It is therefore crucial to design new solutions that support RT and reminiscence conversations without relying on curated personal photographs.
Although RT is often facilitated by healthcare providers such as nurses, in practice it is frequently delivered at home by lay caregivers, such as spouses and adult children, who interact with patients more often but have no formal healthcare training. This lay caregiver-facilitated practice is often referred to as informal RT~\cite{Woods2018Reminiscence}.
Beyond the challenges of curating personal photographs, informal RT facilitated by lay caregivers also face communication challenges~\cite{Pond2026Training}; caregivers may lack the knowledge or shared experiences with the patient needed to facilitate an engaging reminiscence conversation.

Many recent digital technologies have integrated pretrained Large Language Models~(LLMs) to support RT and reminiscence conversations, using imported personal photographs~\cite{Remme2026}, images co-created by the patients and AI~\cite{Zhang2025RemiHaven, Sun2025Chorus, RemiVoice2026}, or a combination of both~\cite{Nan2025Kimono}.
However, most of these technologies are designed to support dyadic conversations between the patient and the AI.
Although lay caregivers such as spouses and adult children are often advised to accompany older adults during AI-facilitated reminiscence conversations to prevent emotional flooding and unexpected distress~\cite{RemiVoice2026}, existing AI-enabled RT technologies often overlook the role of lay caregivers.
While a few recent studies, such as RemiAssist~\cite{Xu2026RemiAssist}, have introduced tools that help caregivers facilitate reminiscence conversations, these tools still require personal photographs.

In this paper, we present our design and early-stage exploration of \emph{MemoryLane}, an AI-integrated tool that supports lay caregivers in facilitating reminiscence conversations and informal RT with their patient with mild dementia.
MemoryLane consists of a smartphone-based interface visible only to the lay caregiver (Figure~\ref{fig::teaser}b) and a shared interface, displayed on a tablet or screen, that is visible to both the patient and their lay caregiver (Figure~\ref{fig::teaser}c).
Rather than facilitating dyadic interactions between a patient with mild dementia and AI, MemoryLane supports triadic reminiscence conversations among the patient, their lay caregiver, and a simulated AI agent (Figure~\ref{fig::teaser}a).
The simulated AI agent, displayed on MemoryLane’s shared interface, acts as a peer to the patient with mild dementia.
This design aims to encourage the patient to share memories more readily, since people tend to open up to others with similar cultural backgrounds or life experiences~\cite{Park2019RTMeta, Zhong2023LifeReview, Yanagida2024GroupRT}.
Although AI-generated content may trigger unexpected reactions such as emotional flooding, distress, or trauma-related responses~\cite{li2026llm, khan2022reminiscence}, MemoryLane adopts a caregiver-in-the-loop approach to keep reminiscence conversations safe. More specifically, the lay caregiver can use MemoryLane's smartphone-based interface to preview and moderate AI-generated images, as well as control what the simulated AI agent says before it joins the triadic conversation (Figure~\ref{fig::teaser}b).
Through a preliminary focus group with the members of the Dementia Care and Cure Initiative (DCCI) Miami Task Force~\cite{DCCI}, we demonstrated the feasibility and value of the simulated AI agent as a peer companion, the value of AI-generated images as memory cues, and the feasibility of having caregivers moderate images and control the simulated AI agent.

\section{MemoryLane}

MemoryLane supports informal RT between a person with mild dementia and their lay caregiver by enabling a triadic reminiscence conversation among the patient, the caregiver, and an AI agent.
The AI agent simulates a peer who shares the patient's cultural background and life experiences.
This is achieved through a dedicated persona that the caregiver pre-configures with key information about the patient, such as their cultural background and significant life experiences.

MemoryLane includes two interfaces: a shared interface visible to both the patient and their caregiver, and a private smartphone interface visible only to the caregiver.
The shared interface, rendered on a shared display or tablet, shows the AI agent's speech transcript alongside AI-generated images created and/or edited on the fly, with the caregiver moderating and controlling the content.
The private smartphone interface that is only visible to caregivers allows them to control the behavior of the simulated AI agent and preview AI-generated images before they appear on the shared display.
While prior work has identified risks in using AI-generated images for RT, as they can unexpectedly trigger reactions such as emotional flooding and distress~\cite{RemiVoice2026}, MemoryLane keeps the caregiver in the loop to prevent such reactions.
The caregiver's interactions with the private smartphone interface are designed to be glanceable and quick, so that controlling and moderating the AI agent does not disrupt the triadic reminiscence conversation among the patient, the caregiver, and the simulated AI agent. 
The private smartphone interface lets the caregiver control the AI agent's behavior and moderate AI-generated images through three mechanisms:
{\bf (1)~Deciding when the AI agent joins.} The simulated AI agent listens to and transcribes the reminiscence conversation between the patient and their lay caregiver, but only intervenes in the conversation when invited by the lay caregiver.
{\bf (2)~Deciding what the AI agent says.} As the patient and caregiver converse, MemoryLane generates candidate topics the AI agent could raise and displays them on the caregiver's private smartphone interface. Selecting a topic results in the agent joining the conversation.
{\bf (3)~Deciding when images are generated or edited.} MemoryLane allows the caregiver to request photorealistic AI-generated images based on the conversation context. Each image is first shown on the caregiver's smartphone for review and, once approved, appears on the shared display.
These AI-generated images serve as the trigger point for the reminiscence conversation among the patient, their lay caregiver, and the simulated AI agent.
Using the smartphone interface, caregivers can continuously request edits to the images as the reminiscence conversation evolves.
Instead of requiring caregivers to write text-to-image prompts manually, MemoryLane uses a dedicated agentic pipeline to generate prompts from the transcribed and diarized conversation.

\textbf{Implementation.} MemoryLane is implemented as a Python-based web application. We use Google Cloud Speech-to-Text~\cite{GoogleSTT} for transcription and diarization, Gemini 3.5 Flash-Lite~\cite{GoogleGemini} as the LLM, Gemini 2.5 Flash Image~\cite{GeminiFlashImage} for image generation and editing, and Gemini 2.5 Flash-Lite TTS~\cite{GoogleGemini} for speech synthesis, configured to produce an empathetic and caring voice.

\section{Preliminary Focus Group}

To gather early feedback on the feasibility of MemoryLane, we conducted an exploratory focus group with the DCCI Miami Task Force~\cite{DCCI}, a community group working in dementia care in South Florida. This study was IRB-approved.

\textbf{Participants and procedures.} We recruited five stakeholders, including an entrepreneur of a dementia-care startup (P1), a family caregiver (P2), a memory-care specialist (P3), a dementia-care volunteer (P4), and a healthcare administrator (P5). All had at least one year of experience in dementia care. Four were affiliated with nonprofit or community organizations, and one with a public health organization. 
After giving informed consent, participants completed a demographic questionnaire about their experience with AI and dementia care. They then watched a video demonstrating MemoryLane's main features, followed by a discussion facilitated by three researchers with backgrounds in computing, geriatrics, and nursing. The discussion covered the affordances of the AI agent and AI-generated images, as well as the feasibility of a private smartphone interface for the caregiver. The qualitative data were analyzed using thematic analysis~\cite{Braun2006Thematic} with both deductive and inductive coding.

\textbf{Findings.} 
We summarize our results in three parts.
\textbf{(1) Value of the simulated AI agent.} All participants saw value in the simulated AI agent that acts as a peer, especially for lay caregivers without formal training. P4 and P5 felt the system would suit patients in the mild stage of dementia. Drawing on his experience reminiscing with his wife, P2 noted that people with dementia ``\textit{will connect with people, but connecting with a screen can be challenging}'' and that a relatable companion with a shared background could help keep the patient focused. He suggested that the simulated AI agent use the patient's name more often to draw their attention back.
\textbf{(2) AI-generated images.} Participants found the pictures useful for prompting memories, but raised three concerns. First, P2 noted that the strongest memories might still come from the patient's own photos, recalling how a picture of his wife's dog instantly brought back her childhood. 
Second, P3 worried that places change over time, so an image that does not match how a building looked back then could cause a disconnect~\cite{Nan2025Kimono}; MemoryLane’s editing feature addresses this issue by drawing on the real-time conversation context.
Third, P2 pointed out that patients' short attention spans make AI inference speed important. Participants also valued the gallery: P4 described how a caregiver could later invite the patient to revisit the pictures from an earlier conversation, making it ``\textit{a real person-centered interactive experience.}''
\textbf{(3) Feasibility of the caregiver's smartphone interface.} Participants considered it feasible for a caregiver to run the system while facilitating the conversation. However, P1 noted that many family caregivers are spouses aged 65 to 80, who may need training, and suggested that home health aides, who are often younger, could be well suited to the role.

\section{Future Work}

Our future work has three directions. First, MemoryLane's current design does not include a dedicated architecture for managing long-term memory across reminiscence conversations. Since participants noted that reminiscence conversations often last 30 minutes or longer, we plan to develop a dedicated memory architecture to support extended sessions. Second, our focus group revealed that many lay caregivers are themselves older adults who may not be tech-savvy. We will therefore explore simplified versions of the smartphone interface that caregivers use to control the AI agent and moderate AI-generated content. Finally, this poster presents only our exploratory focus group with members of the DCCI Miami Task Force. In the long term, we plan to conduct a trial with people with mild dementia and their caregivers to evaluate MemoryLane's usability, utility, and health-related effectiveness.

\balance
\bibliographystyle{ACM-Reference-Format}
\bibliography{reference}

\end{document}